\documentclass[twocolumn,showpacs,preprintnumbers,aps,prl,amsmath,amssymb]{revtex4}
\usepackage{graphicx}
\usepackage{dcolumn}
\usepackage{bm}
\begin{document}
\title{Control of spontaneous emission of an inverted Y-type atomic system coupled by three coherent fields}
\author{Jianbing Qi}
\affiliation{Department of Physics and Astronomy, Penn State University, Berks Campus, Tulpehocken Road, P.O. Box 7009, Reading, PA 19610}

\begin{abstract}
We investigate the spontaneous emission from an inverted Y-type atomic system coupled by three coherent fields. We use the Schr\"{o}dinger equation to calculate the probability amplitudes of the wave function of the system and derive an analytical expression of the spontaneous emission spectrum to trace the origin of the spectral features. Quantum interference effects, such as the spectral line narrowing, spectrum splitting and dark resonance are observed. The number of spectral components, the spectral linewidth, and relative heights can be very different depending on the physical parameters. A variety of spontaneous emission spectral features can be controlled by the amplitudes of the coupling fields and the preparation of the initial quantum state of the atom. We propose an ultracold atomic $^{87}Rb$ system for experimental observation.
\end{abstract}
\pacs{42.50.Gy, 42.50.Ct, 32.80.Qk}


\maketitle
\section{I. Introduction}
It is well-known that the spontaneous emission results from the inevitable interaction of the atomic system with the quantized electromagnetic field vacuum~\cite{QuantumOptics}. Interference effect in the spontaneous emission has become an important topic and the control of the spontaneous emission has attracted intensive study in many atomic systems in recent years~\cite{JPB.15.65.1982,PRL81293Paspalakis,PRL.91.233601.2003}. The process of vacuum fields driving the excited atom to its decay target state can be altered by coupling the excited state to other internal atomic states or by modifying the vacuum states~\cite{PRL47.233.1981}. The spontaneous emission of the atom will be influenced as either the excited state or the decay target state being coupled to an internal state by a coherent field. It is well known that the spontaneous emission spectrum of a two-level atom driven by a strong resonant field is greatly modified~\cite{Mollow triplets,PRL.35.1426.1975}.

The early work of Agarwal has showed that the spontaneous emission of a V-type three-level atom can be modified through the quantum interference between the two decay channels with a common ground level~\cite{Agarwal1974}. The competition among the multiple decay path ways to a common state can result in a complete destructive quantum interference as well as constructive quantum interference~\cite{PRA.56.3011.1997}. It has been shown that the spontaneously generated atomic coherence exists when two close-lying excited atomic levels are coupled by the same vacuum modes to a single lower level with non-orthogonal electric dipole matrix elements between the upper pair and lower states~\cite{PRL81293Paspalakis,PRL76.388.1996Zhu}. However, there are few atomic systems that satisfy this condition, therefore few experiments have been reported~\cite{PRL77.1032.1996Xia,PRL84.4016.2000}. It is natural to think about using coherent fields to couple the involved atomic states to make the atomic system evolving in a controllable way instead of seeking atomic systems with limited operational parameters. Since the atomic states are dressed by the coherent fields, the dressed states will evolve complete differently in the vacuum. For example, the spectral linewidth which is associated with the  decay of one of the dressed states of the atoms depends crucially on the relative strength of the coupling fields and the phase and can be extremely narrowed~\cite{PRL81293Paspalakis}. Understanding the dynamics of the controlled system has many potential applications, such as lasing without inversion, coherent population trapping(CPT)~\cite{EArimondo.1996}, electromagnetically induced transparency(EIT)~\cite{PhysToday50.736.1997}, and fluorescence quenching~\cite{PRA55.4454.1997}.

A variety of atomic coherence and quantum interference phenomena have been discovered in many atomic and molecular systems based on two or three energy state models~\cite{Sci.263.337.1994,JCP104.7068.1996,PRA69.023401.2004,PRA58.4116.1998}, such as coherent population trapping (CPT) ~\cite{PRA53.1014.1995, PRA53.r27.1996}, electromagnetic induced transparency (EIT) ~\cite{PRL74.2431.1995,OPC138.185.1997,PRL84.5094.2000,PRA.73.043810.2006} ultraslow propagation of light ~\cite{Nature397.594.1999,PRL82.5229.1999c}, and Autler-Townes splitting ~\cite{PRA60.450.1999,JCP114.276.2001,PRA71.023401.2005}. Multilevel atomic and molecular systems offer many possibilities for the investigation of coherence effects and quantum control of the interactions among the quantum participants. The multilevel quantum systems provide rich coupling schemes and thus degrees of freedom of controlling parameters. However, the cost of rich coupling configurations will introduce more complexity and difficulty in the experiments and in the theoretical analysis, as well. The experimental realization of a theoretical model is very important to test the understanding of the model. Particularly, the Doppler effect is a severe limit in the observation of many coherence effects~\cite{QIAT99,PRA.73.043810.2006}. Recently, EIT in ultracold atomic gases, and Autler-Townes splitting effect in ultracold molecule formation and detection have been reported~\cite{PRA56.2221.1997,PRA64.061401.2001,PRA68.051403.2003}. The optical information can be coherently controlled with matter wave dynamics in Bose Einstein condensates~\cite{nature445.623.2007}. The development of ultracold physics makes the observation of some subtle coherence effects possible in Doppler free environments otherwise difficult or even impossible at high temperatures.

In this paper, we study the spontaneous emission of a coherently driven inverted-Y type atom as illustrated schematically in Fig. 1. This scheme can be applied to the atomic $^{87}Rb$ system and the corresponding energy levels can be chosen as shown in the parentheses of Fig. 1. We propose an ultracold atomic $^{87}Rb$ to be used to observe the phenomena discussed in the following experimentally in which the Doppler effect is negligible. The ultracold atomic sample can be obtained using a magneto-optical trap(MOT). Similar schemes have been used for the study of Autler-Townes effect in a sodium dimmer ~\cite{JCP124.084308.2006} and the suppression of two-photon absorption ~\cite{PRA64.043807.2001}.

The primary interest of this work is to investigate the controllability of the spectral features of the spontaneous emission of the atom by the coupling fields and other parameters of the system. We use the wave function approach in this paper to obtain an explicit expression for the spontaneous emission spectrum. We find that the spectral features depend upon the amplitude of the coupling fields and the initial quantum state of the atom being prepared. Quantum interference, such as the spectral lineshape narrowing, and fluorescence quenching is observed. We show that the spontaneous emission spectral features can be controlled by the amplitude and detuning of three coupling lasers. We provide a numerical and qualitative analysis to trace the origin of the spectral features, which are attributed to the quantum interference due to the existence of competitive pathways generated by the coherent couplings.

The paper is organized as follows. In section II, we present the description of the model and the derivation of an analytical expression of the spontaneous emission spectrum for the proposed atomic system. We discuss the spectral features and the corresponding numerical simulation in section III, and a summary of our results and some conclusions are given in section IV.
\begin{figure}
\centering \vskip -4 mm
\includegraphics[width=8 cm]{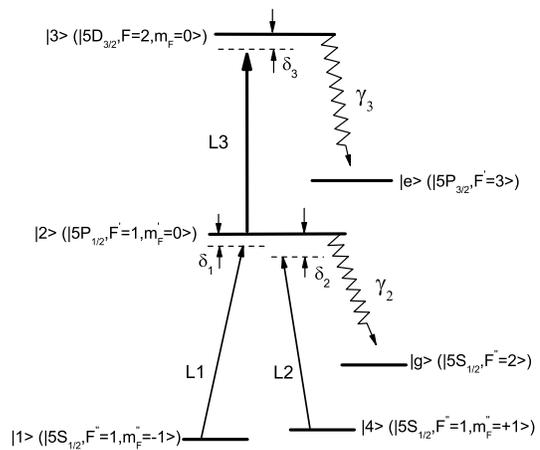}
\caption{The energy level diagram and the coupling scheme. The proposed corresponding atomic $^{87}Rb$ levels are shown in the parentheses.}
\end{figure}
\maketitle
\section{II. Description of the Model and Equations of Motion}
We consider an inverted Y-type atomic system coupled by three lasers as shown in Fig. 1. Laser L1 and L2 couple the ground state level $|1\rangle$ and $|4\rangle$ to a common excited state level $|2\rangle$, respectively, which forms a widely used $\Lambda$ three-level system. In addition, a third laser L3 couples the intermediate excited state level $|2\rangle$ to a higher excited state level $|3\rangle$. Except for the decay of the excited state $|2\rangle$ back to ground state levels $|1\rangle$ and $|4\rangle$, and the upper state $|3\rangle$ to the intermediate state $|2\rangle$, respectively, we assume that the intermediate level $|2\rangle$ decays also to another ground state $|g\rangle$ and the upper excited state $|3\rangle$ decays to an auxiliary intermediate level $|e\rangle$. Both transitions are assumed to be coupled by the vacuum modes in free space. If the separation between level $|2\rangle$ and level $|g\rangle$ is much different from that of level $|3\rangle$ from level $|e\rangle$, we can assume that the vacuum modes coupling between $|2\rangle$ and $|g\rangle$ is totally different from that between $|3\rangle$ and $|e\rangle$. This is true in our scheme since two transition frequencies are very different from each other. The intermediate excited level $|2\rangle$ and the upper excited level $|3\rangle$ have opposite symmetry, and Level $|e\rangle$ has the same symmetry as level $|2\rangle$. The interactions between the coupling fields and the vacuum modes are neglected here. Under the electric dipole and rotating-wave approximation the Hamiltonian of the system in the Schr\"{o}dinger's picture can be written as:
\begin{eqnarray}
H = H_{0}+H_{int}^{s},
\label{eq:one}
\end{eqnarray}
where $H_{0}$ is
\begin{eqnarray}
H_{0}&=&\sum_{i=1}^{4}\hbar\omega_{i}|i\rangle\langle{i}|
       +\hbar\omega_{e}|e\rangle\langle{e}| + \hbar\omega_{g}|g\rangle\langle{g}|\nonumber\\
      & & +\sum_{k}\hbar\omega_{k}b_{k}^{\dag}b_{k}
        +\sum_{q}\hbar\omega_{q}b_{q}^{\dag}b_{q},
\end{eqnarray} and $H_{int}^{s}$ is the interaction Hamiltonian in the Shr\"{o}dinger picture, which is
\label{eq:2}
\begin{eqnarray}
H_{int}^{s}&=&-\frac{\hbar}{2}(\Omega_{12}e^{-i\nu_{1}t}|1\rangle\langle{2}|
    +\Omega_{24}e^{-i\nu_{2}t}|2\rangle\langle{3}|\nonumber\\
    &&+\Omega_{23}e^{-i\nu_{3}t}|4\rangle\langle{2}|)
    +\hbar(\sum_{k}g_{k}b_{k}|g\rangle\langle{2}|\nonumber\\
    &&+\sum_{q}g_{q}b_{q}|e\rangle\langle{3}|)+h.c..
\end{eqnarray}
\label{eq:3}
The $\hbar\omega_{i}$ ($\emph{i}=1, 2,\ldots$) is the energy of the state $|i\rangle$, $\nu_{i}$ ($\emph{i}=1,2,3$) is the laser frequency with the corresponding Rabi frequency defined as $\Omega_{12}=\frac{\mu_{12}\cdot{E_{1}}}{\hbar}$, $\Omega_{24}=\frac{\mu_{24}\cdot{E_{2}}}{\hbar}$, and  $\Omega_{32}=\frac{\mu_{32}\cdot{E_{3}}}{\hbar}$. $\mu_{ij}$ is the electric dipole transition moment of $|i\rangle\leftrightarrow|j\rangle$ transition and $E_{i}$ is the field amplitude of the corresponding coupling laser. $b_{k}^{\dagger}(b_{q}^{\dagger})$ and $b_{k}(b_{q})$ are the photon creation and annihilation operators, and the index k(q) stands for the $kth(qth)$ field mode with frequency $\omega_{k}(\omega_{q})$. The $g_{k}$($g_{q}$) stands for the vacuum coupling constant between the $kth(qth)$ vacuum mode and the atomic transitions $|2\rangle\leftrightarrow|g\rangle$($|3\rangle\leftrightarrow|e\rangle$). The summation over k(q) runs over modes near the corresponding atomic transition. For simplicity of the calculation but without loss of the generality, we take the energy of level $|1\rangle$ as the reference, and let $\omega_{1}=0$. The Hamiltonian in the interaction representation is obtained through the following transformation:
\begin{eqnarray}
H_{I} = e^{iH_{0}t/\hbar}H_{int}^{s}e^{-iH_{0}t/\hbar},
\end{eqnarray}
\label{eq:4}
which is can be explicitly written as,
\begin{eqnarray}
H_{I}(t)&=&-\frac{\hbar}{2}(\Omega_{12}e^{-i\delta_{1}t}|2\rangle\langle{1}|
    +\Omega_{24}e^{-i\delta_{2}t}|3\rangle\langle{2}|\nonumber\\
    &&
    +\Omega_{32}e^{-i\delta_{3}t}|2\rangle\langle{4}|)
    +\hbar(\sum_{k}g_{k}e^{-i\delta_{k}t}b_{k}|2\rangle\langle{g}|\nonumber\\
    &&
    +\sum_{q}g_{q}e^{-i\delta_{q}t}b_{q}|3\rangle\langle{e}|)+h.c.,
\end{eqnarray}
\label{eq:5}
where $\delta_{1}=\nu_{1}-\omega_{21}$, $\delta_{2}=\nu_{2}-\omega_{24}$, and $\delta_{3}=\nu_{3}-\omega_{32}$ are the frequency detunings of laser L1, L2, and L3, respectively; $\delta_{k}=\omega_{k}-\omega_{2g}$ and $\delta_{q}=\omega_{q}-\omega_{3e}$ are the frequency detunings of the spontaneous emission with respect to the transition, $|2\rangle\rightarrow|e\rangle$, and $|3\rangle\rightarrow|g\rangle$, respectively. The state vector of the system at any time t can be expanded in terms of bare-state eigenvectors of the system as
\begin{eqnarray}
|\Psi(t)\rangle=&&[a_{1}(t)|1\rangle
                    + a_{2}(t)|2\rangle
                    + a_{3}(t)|3\rangle
                    + a_{4}(t)|4\rangle]|\{0\}\rangle\nonumber\\
                & &
                    +\Sigma_{k}
                    a_{g,k}(t)|g\rangle|1_{k}\rangle
                    +\Sigma_{q}
                    a_{e,q}(t)|e\rangle|1_{q}\rangle,
\end{eqnarray}
\label{eq:6}
where $|i\rangle$ (\textit{i}=1,2...4) is the \textit{i}th unperturbed stationary state of the atom, $|\{0\}\rangle$ represents for the absence of photons in all vacuum modes of the field, and $|1_{k}\rangle(|1_{q}\rangle)$ denotes that there is one photon in the $kth(qth)$ vacuum mode. The $a_{i}(t)'s$, $a_{g,k}$, and $a_{e,q}$ are the probability amplitudes for the atomic state $|i\rangle, |g\rangle$, and $|e\rangle$, respectively. The initial values of the corresponding probability amplitudes of the state vector depend upon the initial preparation of the atom. We assume that $a_{g,k}(0)=a_{e,q}(0)=0$,  and $a_{i}(0)'s$ are arbitrary, apart from the normalization requirement, $\sum_{i=1}^{4}|a_{i}(0)|^2=1$. Then the Schr\"{o}dinger equation in the interaction picture is
\begin{eqnarray}
\frac{\partial{|\Psi(t)\rangle}}{\partial{t}}=-\frac{i}{\hbar}H_{I}|\Psi(t)\rangle.
\end{eqnarray}
\label{eq:7}
The equations of motion for the probability amplitudes of the wave function are readily obtained by substituting Eq.~(5) and Eq.~(6) into Eq.~(7) and applying the Weisskopf-Wigner theory.
\begin{subequations}
\label{eq:whole 8a-8f}
\begin{eqnarray}
\dot{a}_{1}(t) = i\frac{\Omega_{12}^{*}}{2}e^{i\delta_{1}t}a_{2}(t)
\end{eqnarray}
\begin{eqnarray}
\dot{a}_{2}(t)&=&
                i\frac{\Omega_{12}}{2}e^{-i\delta_{1}t}a_{1}
                +i\frac{\Omega_{24}}{2}e^{-i\delta_{2}t}a_{4}(t)\nonumber\\
                &&
                +i\frac{\Omega_{23}^{*}}{2}e^{i\delta_{3}t}a_{3}(t)
                -\frac{\gamma_{2}}{2}a_{2}(t)
\end{eqnarray}
\begin{eqnarray}
\dot{a}_{3}(t) = i\frac{\Omega_{23}}{2}e^{-i\delta_{3}t}a_{2}(t)
                -\frac{\gamma_{3}}{2}a_{3}(t)
\end{eqnarray}
\begin{eqnarray}
\dot{a}_{4}(t) = i\frac{\Omega_{24}^{*}}{2}e^{i\delta_{2}t}a_{2}(t)
\end{eqnarray}
\begin{eqnarray}
\dot{a}_{g,k}(t) = -ig_{k,2g}e^{i\delta_{k}t}a_{2}(t)
\end{eqnarray}
\begin{eqnarray}
\dot{a}_{e,q}(t) = -ig_{q,3e}e^{i\delta_{q}t}a_{3}(t),
\end{eqnarray}
\end{subequations}
where the $\gamma_{2(3)}=2\pi|g_{k(q)}(\omega_{k(q)})|^{2}D(\omega_{k(q)})$ is the spontaneous decay rate from level $|2\rangle(|3\rangle)$ to level $|g\rangle(|e\rangle)$, and the D($\omega_{k(q)}$) is the vacuum mode density at frequency $\omega_{k(q)}$ in the free space. Using following transformation,
\begin{subequations}
\begin{eqnarray}
C_{1}(t)=a_{1}(t)
\end{eqnarray}
\begin{eqnarray}
C_{2}(t)=a_{2}(t)e^{i\delta_{1}t}
\end{eqnarray}
\begin{eqnarray}
C_{3}(t)=a_{3}(t)e^{i(\delta_{1}+\delta_{3})t}
\end{eqnarray}
\begin{eqnarray}
C_{4}(t)=a_{4}(t)e^{i(\delta_{1}-\delta_{2})t},
\end{eqnarray}
\label{eq:9a-9d}
\end{subequations}
we have six coupled first order differential equations:
\begin{subequations}
\label{eq:10a-10f}
\begin{eqnarray}
\dot{C}_{1}(t)=i\frac{\Omega_{12}^{*}}{2}C_{2}(t)
\end{eqnarray}
\begin{eqnarray}
\dot{C}_{2}(t)= i\delta_{1}C_{2}(t)
                -\frac{\gamma_{2}}{2}C_{2}(t)
                +i\frac{\Omega_{12}}{2}C_{1}(t)\nonumber\\
                +i\frac{\Omega_{23}}{2}^{*}C_{3}(t)
                +i\frac{\Omega_{24}}{2}C_{4}(t)
\end{eqnarray}
\begin{eqnarray}
\dot{C}_{3}(t)= i(\delta_{1}+\delta_{3})C_{3}(t)
                -\frac{\gamma_{3}}{2}C_{3}(t)
                +i\frac{\Omega_{23}}{2}C_{2}(t)
\end{eqnarray}
\begin{eqnarray}
\dot{C}_{4}(t)= i(\delta_{1}-\delta_{2})C_{4}(t)
                +i\frac{\Omega_{24}^{*}}{2}C_{2}(t)
\end{eqnarray}
\begin{eqnarray}
\dot{a}_{g,k}(t)=-ig_{k,2g}e^{i(\delta_{k}-\delta_{1})t}C_{2}(t)
\end{eqnarray}
\begin{eqnarray}
\dot{a}_{e,q}(t)=-ig_{q,3e}e^{i(\delta_{q}-\delta_{2}-\delta_{1})t}C_{3}(t).
\end{eqnarray}
\end{subequations}
Using Laplace transformations $\tilde{C}_{j}(s)=L(C_{j}(t))=\int_{0}^{\infty}e^{-st}C_{j}(t)dt$ for equations (10a-d), and integrating equations 10e and 10f, we obtain the following six equations:
\begin{subequations}
\label{eq:11a-f}
\begin{eqnarray}
s\tilde{C}_{1}(s)-C_{1}(0)=i\frac{\Omega_{12}^{*}}{2}
                            \tilde{C}_{2}(s)
\end{eqnarray}
\begin{eqnarray}
s\tilde{C}_{2}(s)-C_{2}(0)&=& i(\delta_{1}+i\frac{\gamma_{2}}{2})
                \tilde{C}_{2}(s)
                +i\frac{\Omega_{12}}{2}
                \tilde{C}_{1}(s)\nonumber\\
            & & +i\frac{\Omega_{23}^{*}}{2}
                \tilde{C}_{3}(s)
                +i\frac{\Omega_{24}}{2}
                \tilde{C}_{4}(s)
\end{eqnarray}
\begin{eqnarray}
s\tilde{C}_{3}(s)-C_{3}(0)&=& i(\delta_{1}+\delta_{3})\tilde{C}_{3}(s)\nonumber\\
                & &-\frac{\gamma_{3}}{2}\tilde{C}_{3}(s)
                +i\frac{\Omega_{23}}{2}\tilde{C}_{2}(s)
\end{eqnarray}
\begin{eqnarray}
s\tilde{C}_{4}(s)-C_{4}(0)= i(\delta_{1}-\delta_{2})\tilde{C}_{4}(s)
                +i\frac{\Omega_{24}^{*}}{2}\tilde{C}_{2}(s)
\end{eqnarray}
\begin{eqnarray}
\tilde{a}_{g,k}(s)=\frac{-ig_{k,2g}\tilde{C}_{2}(s)}{s+i(\delta_{k}-\delta_{1})}
\end{eqnarray}
\begin{eqnarray}
\tilde{a}_{e,q}(s)=\frac{-ig_{q,3e}\tilde{C}_{3}(s)}{s+i(\delta_{q}-\delta_{1}-\delta_{3})},
\end{eqnarray}
\end{subequations}
where the $C_{j}(0)'s$ are the corresponding initial conditions given by $a_{j}(0)$, which indicate how the atom is initially prepared. From equation 11a-d we obtain following results for $\tilde{C}_{2}(s)$ and $\tilde{C}_{3}(s)$,

\label{eq:12a-b}
\begin{subequations}
\begin{widetext}
\begin{eqnarray}
\tilde{C}_{2}(s)=\frac{a_{2}(0)+\frac{i\frac{\Omega_{12}}{2}a_{1}(0)}{s}
                        -\frac{i\frac{\Omega_{23}^{*}}{2}a_{3}(0)}{s-i(\delta_{1}+\delta_{3})+\frac{\gamma_{3}}{2}}
                        +\frac{i\frac{\Omega_{24}}{2}a_{4}(0)}{s-i(\delta_{1}-\delta_{2})}
                        }
                        {s+\frac{\gamma_{2}}{2}-i\delta_{1}+\frac{|\frac{\Omega_{12}}{2}|^{2}}{s}
                            +\frac{|\frac{\Omega_{24}}{2}|^{2}}{s-i(\delta_{1}-\delta_{2})}
                             +\frac{|\frac{\Omega_{23}}{2}|^{2}}{s-i(\delta_{1}+\delta_{3})+\frac{\gamma_{3}}{2}}
                        }\nonumber\\
\end{eqnarray}
\end{widetext}
and
\begin{eqnarray}
\tilde{C}_{3}(s)= \frac{a_{3}(0)+i\frac{\Omega_{23}}{2}\tilde{C}_{2}(s)}
{s-i(\delta_{1}+\delta_{3})+\frac{\gamma_{3}}{2}}
\end{eqnarray}
\end{subequations}
The spontaneous emission spectrum is proportional to the Fourier transformation of the field correlation function.
\label{eq:13}
\begin{eqnarray}
&&\langle\mathbf{E}^{-}(\mathbf{r},t+\tau)\cdot\mathbf{E}^{+}(\mathbf{r},t)\rangle_{t\rightarrow\infty}\nonumber\\
& &=\langle\Psi_{I}(t)|
\mathbf{E}^{-}(\mathbf{r},t+\tau)
\cdot
\mathbf{E}^{+}(\mathbf{r},t)
|\Psi_{I}(t)\rangle_{t\rightarrow\infty}.
\end{eqnarray}
It can be shown that the spontaneous emission spectrum is
$\mathbf{S}_{2}(\omega_{k})=\frac{\gamma_{2}|a_{g,k}(t\rightarrow\infty)|^{2}}{2\pi|g_{k}(\omega_{k})|^{2}}$ for the transition of $|2\rangle\rightarrow|g\rangle$, and $\mathbf{S}_{3}(\omega_{q})=\frac{\gamma_{3}|a_{e,q}(t\rightarrow\infty)|^{2}}{2\pi|g_{q}(\omega_{q})|^{2}}$, for the transition of $|3\rangle\rightarrow|e\rangle$, respectively. Using the final value theorem ~\cite{Finalvaluetheorem} and Eq. 11e-f, we obtain an analytical expression for the spontaneous emission of the intermediate excited level $|2\rangle$
\begin{subequations}
\label{eq:14a-b}
\begin{eqnarray}
S_{2}(\omega_{k})&=&\frac{\gamma_{2}|a_{g,k}(t\rightarrow\infty)|^{2}}{2\pi|g_{k,2g}(\omega_{k})|^{2}}\nonumber\\
        &=&\frac{\gamma_{2}|\tilde{C}_{2}(s=-i(\delta_{k}-\delta_{1}))|^{2}}{2\pi},
\end{eqnarray}
and similarly, the expression for the upper level $|3\rangle$
\begin{eqnarray}
S_{3}(\omega_{q})&=&\frac{\gamma_{3}|a_{e,q}(t\rightarrow\infty)|^{2}}{2\pi|g_{q,3e}(\omega_{q})|^{2}}\nonumber\\
        &=&\frac{\gamma_{3}|\tilde{C}_{3}(s=-i(\delta_{q}-\delta_{1}-\delta_{3}))|^{2}}{2\pi}.
\end{eqnarray}
\end{subequations}
From equation 12a-b, the explicit form of the $|\tilde{C}_{2}(s=-i(\delta_{k}-\delta_{1}))|^2$ is
\begin{subequations}
\label{eq:15a-c}
\begin{widetext}
\begin{eqnarray}
|\tilde{C}_{2}(\delta_{k})|^2=\left|\frac{a_{2}(0)-\frac{\frac{\Omega_{12}}{2}a_{1}(0)}{\delta_{k}-\delta_{1}}
-\frac{\frac{\Omega_{23}^{*}}{2}a_{3}(0)}{\delta_{k}+\delta_{3}+i\frac{\gamma_{3}}{2}}-
\frac{\frac{\Omega_{24}}{2}a_{4}(0)}{\delta_{k}-\delta_{2}}}{
\frac{\gamma_{2}}{2}-i\delta_{k}-\frac{|\frac{\Omega_{12}}{2}|^{2}}{i(\delta_{k}-\delta_{1})}-
\frac{|\frac{\Omega_{24}}{2}|^{2}}{i(\delta_{k}-\delta_{2})}
-\frac{|\frac{\Omega_{23}}{2}|^{2}}{i(\delta_{k}+\delta_{3})-\frac{\gamma_{3}}{2}}}\right|^2,\nonumber\\
\end{eqnarray}
and similarly for $|\tilde{C}_{3}(s=-i(\delta_{q}-\delta_{1}-\delta_{3}))|^2$,
\begin{eqnarray}
|\tilde{C}_{3}(\delta_{q})|^2=
        \frac{|a_{3}(0)(\frac{\gamma_3}{2}+i\delta_{q})
                +i\frac{\Omega_{23}}{2}
                (\frac{\gamma_3}{2}+i\delta_{q})
                \tilde{C}_2(s=-i(\delta_{q}-\delta_{1}-\delta_{3}))|^2}
                {\delta_{q}^2+\frac{\gamma_2^2}{4}},
\end{eqnarray} with
\begin{eqnarray}
\tilde{C}_{2}(s=-i(\delta_{q}-\delta_{1}-\delta_{3}))=
\frac{a_{2}(0)
        -\frac{\frac{\Omega_{12}}{2}a_{1}(0)}{\delta_{q}-\delta_{1}-\delta_{3}}
        +\frac{i\frac{\Omega_{23}^{*}}{2}a_{3}(0)(\frac{\gamma_{3}}{2}+i\delta_{q})}{\delta_{q}^2+\frac{\gamma_{3}^2}{4}}
        -\frac{\frac{\Omega_{24}}{2}a_{4}(0)}{\delta_{q}-\delta_{2}-\delta_{3}}}
        {\frac{\gamma_{2}}{2}-i(\delta_{q}-\delta_{3})
            +\frac{i|\frac{\Omega_{12}}{2}|^{2}}{\delta_{q}-\delta_{1}-\delta_{3}}
            +\frac{i|\frac{\Omega_{24}}{2}|^{2}}{\delta_{q}-\delta_{2}-\delta_{3}}
            +\frac{|\frac{\Omega_{23}}{2}|^{2}(\frac{\gamma_{3}}{2}+i\delta_{q})}{\delta_{q}^2+\frac{\gamma_{3}^2}{4}}}.
\end{eqnarray}
\end{widetext}
\end{subequations}

\section{III. Discussion and Numerical Results}
From Eqs. (15a-c) we can see that the spontaneous emission depends upon the initial probability amplitudes($a_{i}(0)$) of the atom, or the initial quantum state of the atom being prepared, the Rabi frequency and the frequency detuning of the lasers. Even the analytical spectrum expression is succinct, however it is still very complicated and difficult to identify the physical origin of the spectral features. For this reason, we limit our discussion to the resonant coupling situation, that is, all three lasers are on resonance with respect to the corresponding transitions. We hope that the equations can be simplified enough but not prevent us from understanding the essential physics of the system and analyzing the effects of each physical parameter to the spectral features. By inspecting Eq. (15c), if the atom is initially not prepared in level $|3\rangle$ or at any superposition state involving level $|3\rangle$, for the resonant coupling we can have the similar discussion for level $|3\rangle$ as for level $|2\rangle$ . Therefore we focus our discussion to the spontaneous emission spectrum $S_{2}(\omega_{k})$ of the intermediate level $|2\rangle$ in this paper.

In Fig. 1, if there is no laser L3 the system will be a widely studied $\Lambda$ coupling scheme($|1\rangle-|2\rangle-|4\rangle$), while if without the coupling laser L2 the system ($|1\rangle-|2\rangle-|3\rangle$) will be a so called cascade scheme. Both schemes have been discussed in references~\cite{PRA.43.3748.1991, PRA72.023802.2005,JPhysB.38.3815.2005}. The inverted Y-type scheme might be regarded as an extension of the $\Lambda$ scheme, however it shows that introducing the coupling between the excited state $|2\rangle$ and an upper level $|3\rangle$ not only changes the dynamics of the system significantly but also brings additional options for controlling spontaneous emission of level $|2\rangle$.

For the resonant coupling case, $\delta_1=\delta_2=\delta_3=0$, Eq. (15a) reads
\begin{eqnarray}
\tilde{C}_{2}(s)=\left|\frac{a_{2}(0)+\frac{i\frac{\Omega_{12}}{2}a_{1}(0)
                        +i\frac{\Omega_{24}}{2}a_4(0)}{-i\delta_k}+\frac{i\frac{\Omega_{23}^*}{2}a_{3}(0)}{-i\delta_k+\frac{\gamma_3}{2}}}
                        {-i\delta_k+\frac{\gamma_2}{2}
                         +\frac{|\frac{\Omega_{12}}{2}|^2+|\frac{\Omega_{24}}{2}|^2}{-i\delta_k}
                         +\frac{|\frac{\Omega_{23}}{2}|^2}{-i\delta_k+\frac{\gamma_3}{2}}}\right|^2
\end{eqnarray}
Substituting Eq. (16) into Eq. (14a), the spontaneous emission spectrum $S_{2}(\omega_{k})$ can be written as
\label{eq:17}
\begin{widetext}
\begin{eqnarray}
S_2(\omega_k)=\frac{\gamma_2}{2\pi}\left|\frac{(\frac{\Omega_{12}}{2}a_{1}(0)
                                                +\frac{\Omega_{24}}{2}a_{4}(0)
                                                -a_2(0)\delta_{k})(\delta_k+i\frac{\gamma_3}{2})-\frac{\Omega_{23}^*}{2}a_3(0)\delta_k}
                                            {(-i\delta_{k}+\Lambda_1)(-i\delta_{k}+\Lambda_2)(-i\delta_{k}+\Lambda_3)}\right|^2,
\end{eqnarray}
\end{widetext}
where $\Lambda_1$, $\Lambda_2$, and $\Lambda_3$ are the roots of the following cubic equation.
\label{eq:18}
\begin{eqnarray}
&&s^3+s^2(\frac{\gamma_2+\gamma_3}{2})+s(\frac{|\Omega_{12}|^2+|\Omega_{24}|^2+|\Omega_{23}|^2+\gamma_2\gamma_3}{4})\nonumber\\
&&+\frac{\gamma_3}{2}(\frac{|\Omega_{12}|^2+|\Omega_{24}|^2}{4})=0\
\end{eqnarray}
It is trivial to solve the above equation, and the three roots are:
\begin{subequations}
\label{19a-c}
\begin{eqnarray}
\Lambda_1&=&y_{+}+y_{-}-\frac{\gamma_2+\gamma_3}{6},
\end{eqnarray}
\begin{eqnarray}
\Lambda_2=-\frac{(y_{+}+y_{-})}{2}
            -\frac{\gamma_2+\gamma_3}{6}
            +i\frac{\sqrt{3}}{2}(y_{+}-y_{-}),
\end{eqnarray}
\begin{eqnarray}
\Lambda_3=-\frac{(y_{+}+y_{-})}{2}
            -\frac{\gamma_2+\gamma_3}{6}
            +i\frac{\sqrt{3}}{2}(y_{-}-y_{+}),
\end{eqnarray}
\end{subequations}
where
\label{Eq. 20 for y}
\begin{eqnarray}
y_{\pm}=\sqrt[3]{-\frac{q}{2}\pm\sqrt{(\frac{q}{2})^{2}+(\frac{p}{3})^{3}}},
\end{eqnarray}
and
\label{Eq. 21 for p}
\begin{eqnarray}
p=\frac{|\Omega_{12}|^2+|\Omega_{24}|^2+|\Omega_{23}|^2+\gamma_2\gamma_3}{4}-\frac{(\gamma_2+\gamma_3)^2}{12},
\end{eqnarray}
\label{Eq. 22 for q}
\begin{eqnarray}
q&=&-\frac{\gamma_2+\gamma_3}{6}(\frac{|\Omega_{12}|^2+|\Omega_{24}|^2+|\Omega_{23}|^2+\gamma_2\gamma_3}{4})\nonumber\\
   & & +\frac{\gamma_3}{2}(\frac{|\Omega_{12}|^2+|\Omega_{24}|^2}{4})+2(\frac{\gamma_2+\gamma_3}{6})^3.
\end{eqnarray}
Further inspecting the structure of Eq. (17) we can rewrite the denominator of Eq. (17) as following.
\label{eq:23}
\begin{eqnarray}
S_2(\omega_k)\propto\left|\frac{1}
                            {(\delta_{k}+i\Lambda_1)(\delta_{k}+i\Lambda_2)(\delta_{k}+i\Lambda_3)}
                            \right|^2\nonumber\\
                 =\left|\frac{\beta_1}{\delta_{k}+i\Lambda_1}
                                +\frac{\beta_2}{\delta_{k}+i\Lambda_2}
                                +\frac{\beta_3}{\delta_{k}+i\Lambda_3}\right|^2\nonumber,\\
\end{eqnarray}
where the coefficient $\beta_{i}$ can be determined by some simple algebraic calculations.
\label{coefficients of betas eq:24}
\begin{eqnarray}
\beta_i=\frac{(\Lambda_k-\Lambda_j)\epsilon_{kji}}
                {\Lambda_{1}^2(\Lambda_2-\Lambda_3)
                +\Lambda_{2}^2(\Lambda_3-\Lambda_1)
                +\Lambda_{3}^2(\Lambda_1-\Lambda_2)}
\end{eqnarray}
$\epsilon_{kji}$ is the permutation symbol and i=1,2,3. Eq. (23) shows that the spontaneous emission spectrum $S_{2}(\omega_{k})$ is a square of the sum of three complex quantities, therefore the interference effect is inherent. Of course, the interference can be constructive or destructive depending upon the physical parameters which we will discuss in details in the following.
\subsection{(A) Dark Line and Dark States}
We assume that both $|2\rangle$ and $|3\rangle$ have the same decay rate in our following discussions. By inspecting the numerator of Eq. (17), if the atom is initially prepared in the excited state $|2\rangle$ ($a_2(0)=1$) or $|3\rangle$ ($a_3(0)=1$) or in a superposition state of $|2\rangle$ and $|3\rangle$($\psi(0)=a_2(0)|2\rangle+a_3(0)|3\rangle$), but not in a dark state, we find that the spectrum of $S_2(\omega_k)$ has a complete dark line($S_2(\omega_k)=0$) due to a destructive interference at the resonant frequency $\delta_k=0$. The spectrum always has a dark line at the resonance frequency and two components as long as the atom is initially prepared in the excited states as shown in Fig. 2(a)-(c). This result is similar to the earlier work for a cascade three level system by Zhu et al.~\cite{PRA.52.4791.1995}.
\begin{figure}
\centering
\includegraphics[width=8 cm]{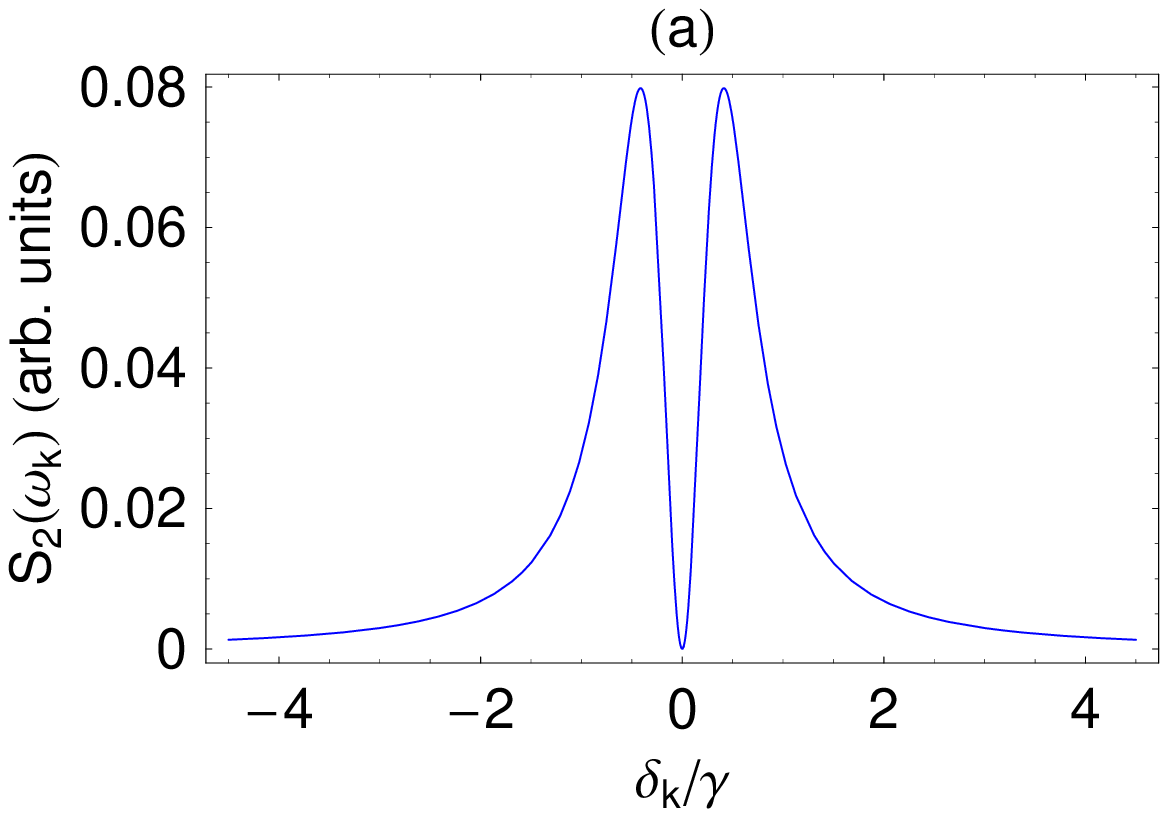}
\includegraphics[width=8 cm]{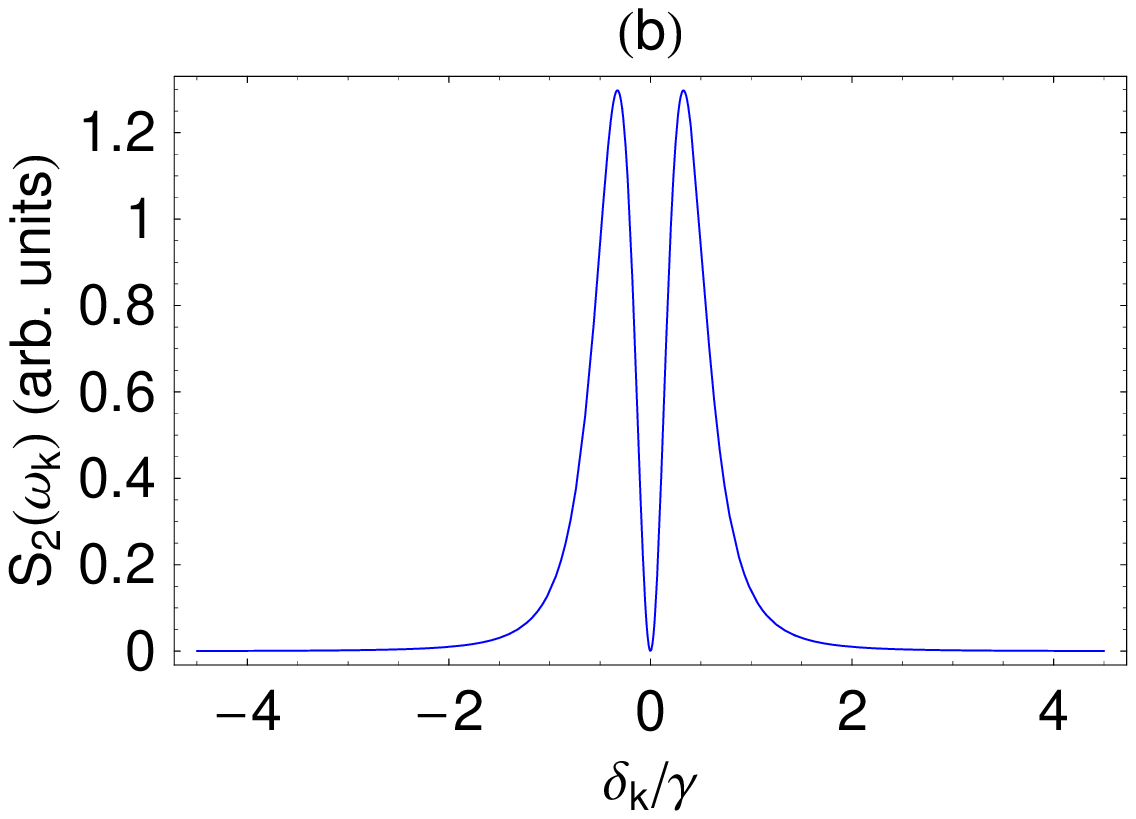}
\includegraphics[width=8 cm]{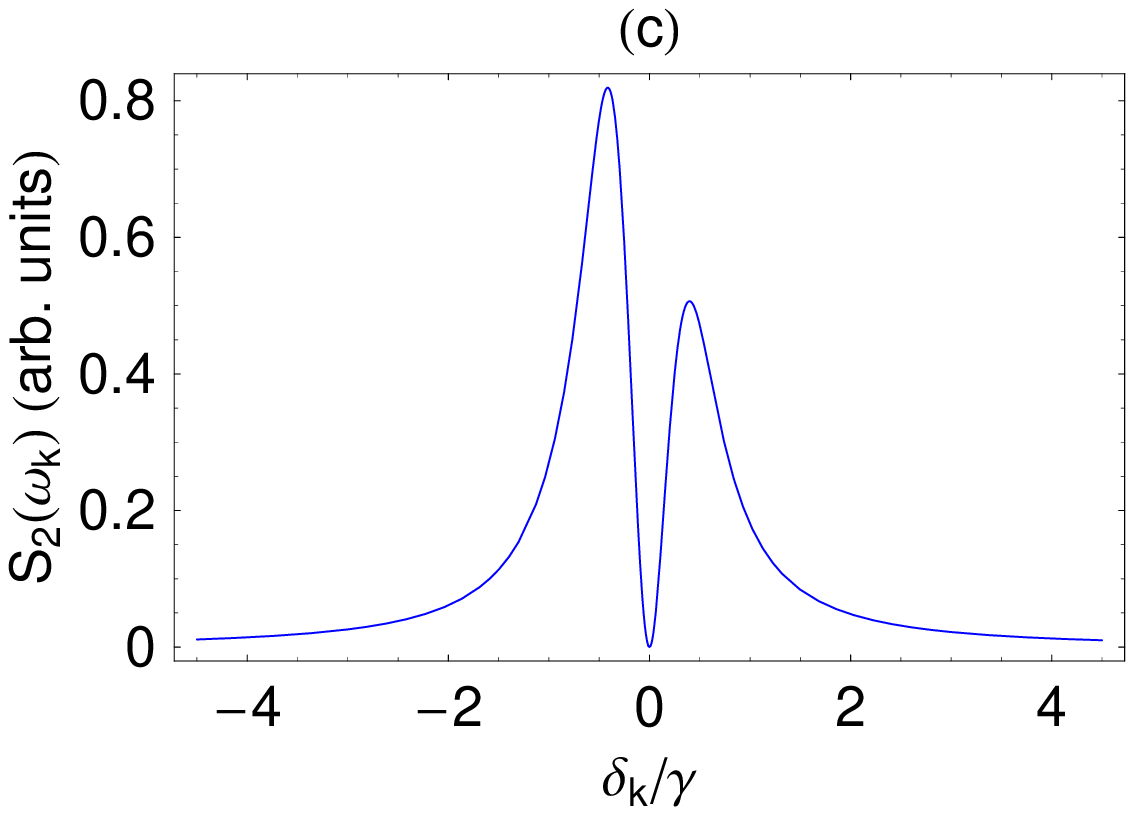}
\includegraphics[width=8 cm]{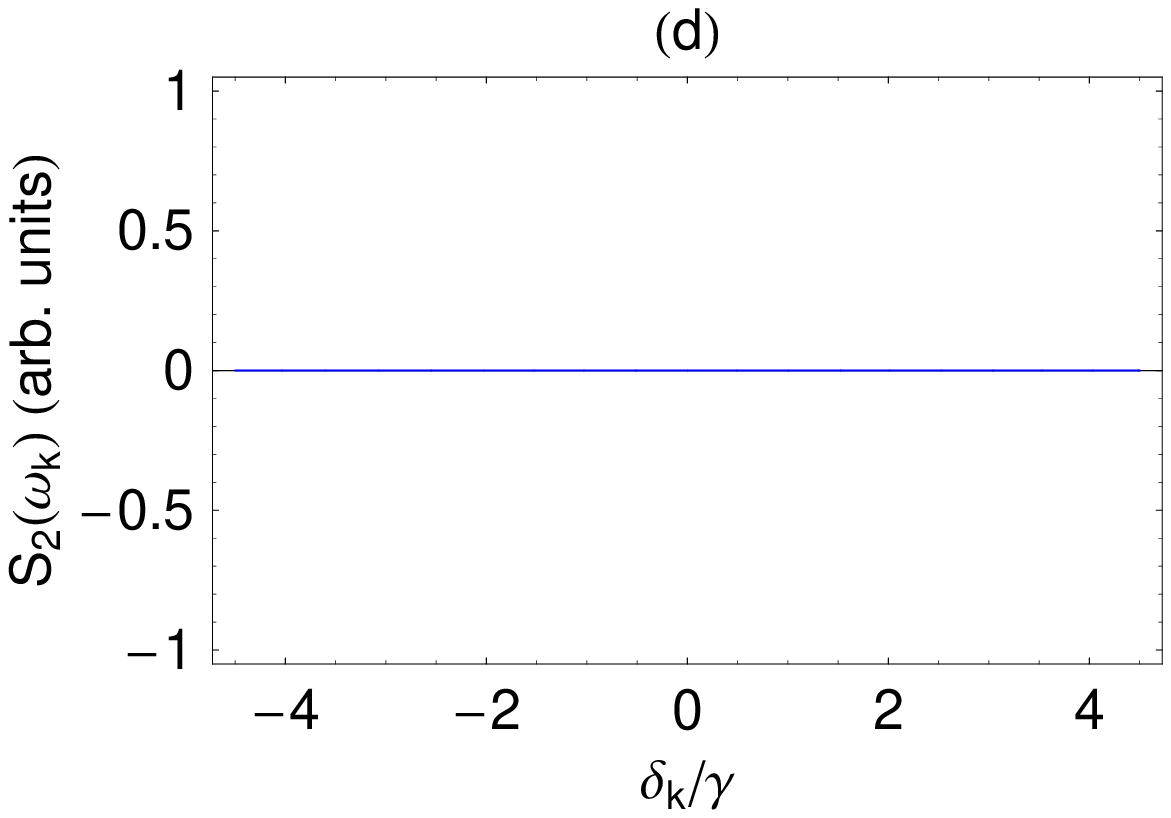}
\caption{(Color online) The spontaneous emission spectra $S_2(\omega_k)$ of the intermediate state
$|2\rangle$. The parameters for the calculations are $\Omega_{12}=\Omega_{24}=\Omega_{23}=0.5\gamma$,
$\delta_{1}=\delta_{2}=\delta_{3}=0$,
$\gamma_{2}=\gamma_{3}=\gamma$, and $\gamma=6.0$ MHz.
(a) $a_{1}(0)=a_{4}(0)=a_3(0)=0$,$a_2(0)=1$, (b) $a_{1}(0)=a_{4}(0)=a_2(0)=0$, $a_3(0)=1$
and (c) $a_{1}(0)=a_{4}(0)=0$, $a_{2}(0)=\sqrt{0.8},a_{3}(0)=\sqrt{0.2}$.
(d)The atom is in a dark state: $a_1(0)=-a_4(0)=\sqrt{0.5}$, and $a_2(0)=a_3(0)=0$.}
\end{figure}
We note that there can be a dark state, in which the atom will be completely decoupled from the interaction of laser 3 and stay at the dark state. As we can see in Eq. (17) that if the atom is initially not prepared in the excited states, $a_3(0)=a_2(0)=0$, but in a superposition of two ground states: $|\psi(0)\rangle=a_1(0)|1\rangle+a_4(0)|4\rangle$,  and the Rabi frequencies of laser L1 and L2 are chosen such that $\frac{\Omega_{12}}{2}a_{1}(0)+\frac{\Omega_{24}}{2}a_{4}(0)=0$, then the atom stays in a dark state and the spontaneous emission of the atom will be completely suppressed. One such example is illustrated in Fig. 2(d).
\subsection{(B) Spectral Line Splitting and Narrowing}
From an experiment perspective, the atom is typically prepared in a single ground state at start, such as in level $|1\rangle$(that is, $a_1(0)=1$) or in level $|4\rangle$(that is, $a_4(0)=1$). Of course, $S_2(\omega_k)$ should have a similar spectral lineshape for both cases. This can readily be checked in Eq. (17). Without the loss of generality, we assume the atom is initially prepared in level $|1\rangle$ with $a_1(0)=1$. Substituting Eq. (19a-c) into Eq. (17) we obtain
\label{eq:26}
\begin{eqnarray}
S_2(\omega_k)=\frac{\frac{\gamma_2}{2\pi}|\frac{\Omega_{12}}{2}|^2(\delta_k^2+\frac{\gamma_3^2}{4})}
                    {[\delta_{k}^2+\Gamma_1^2]
                    [(\delta_{k}-\delta_{\lambda})^2+\Gamma_2^2]
                    [(\delta_{k}+\delta_{\lambda})^2+\Gamma_2^2]},\nonumber\\
\end{eqnarray}
where
\label{eq 27a-c: linewidth and position}
\begin{subequations}
\begin{eqnarray}
\delta_{\lambda}&=&Im\Lambda_2=-Im\Lambda_3=\frac{\sqrt{3}}{2}(y_{+}-y_{-}),\\
\Gamma_1&=&\Lambda_1=y_{+}+y_{-}-\frac{\gamma_2+\gamma_3}{6},\\
\Gamma_2&=&Re\Lambda_2=Re\Lambda_3=-\frac{(y_{+}+y_{-})}{2}-\frac{\gamma_2+\gamma_3}{6}.\nonumber\\
\end{eqnarray}
\end{subequations}
Eq. (25) indicates that the spectrum of $S_2(\omega_k)$ has three peaks, one at resonance frequency $\delta_k=0$ with a spectral width of $2|\Gamma_1|$, and two symmetric sidebands at $\delta_k=\pm\delta_{\lambda}$ with a spectral width of $2|\Gamma_2|$, respectively. The linewidth of the two sidebands is always larger than that of the central component. Both the linewidth and the position of the sidebands depend upon the Rabi frequencies of lasers. However, the spectral features can be significantly different for various combinations of Rabi frequencies of three lasers as we will see in the following numerical calculations. We scale the decay rates, Rabi frequencies and the frequency of the spontaneous emission by the decay rate of level $|2\rangle$ in our calculations.

There is no surprise that the spectrum shows just a single resonance peak when the Rabi frequency of the lasers is smaller than the decay rate of level $|2\rangle$(as shown in Fig. 3(a)). As the Rabi frequencies of lasers increase to the decay rate of level $|2\rangle$ the spectrum starts to display some broadening structures (Fig. 3(b)).
\begin{figure}
\centering
\includegraphics[width=4 cm]{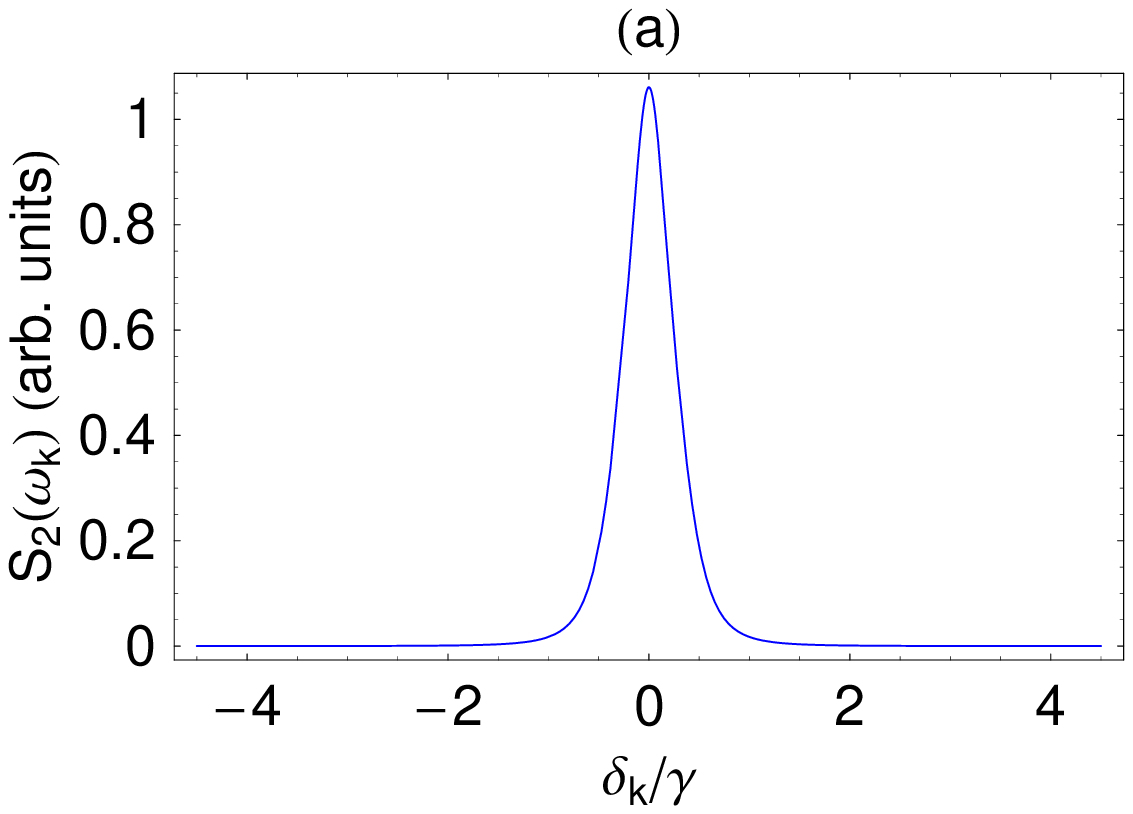}
\includegraphics[width=4 cm]{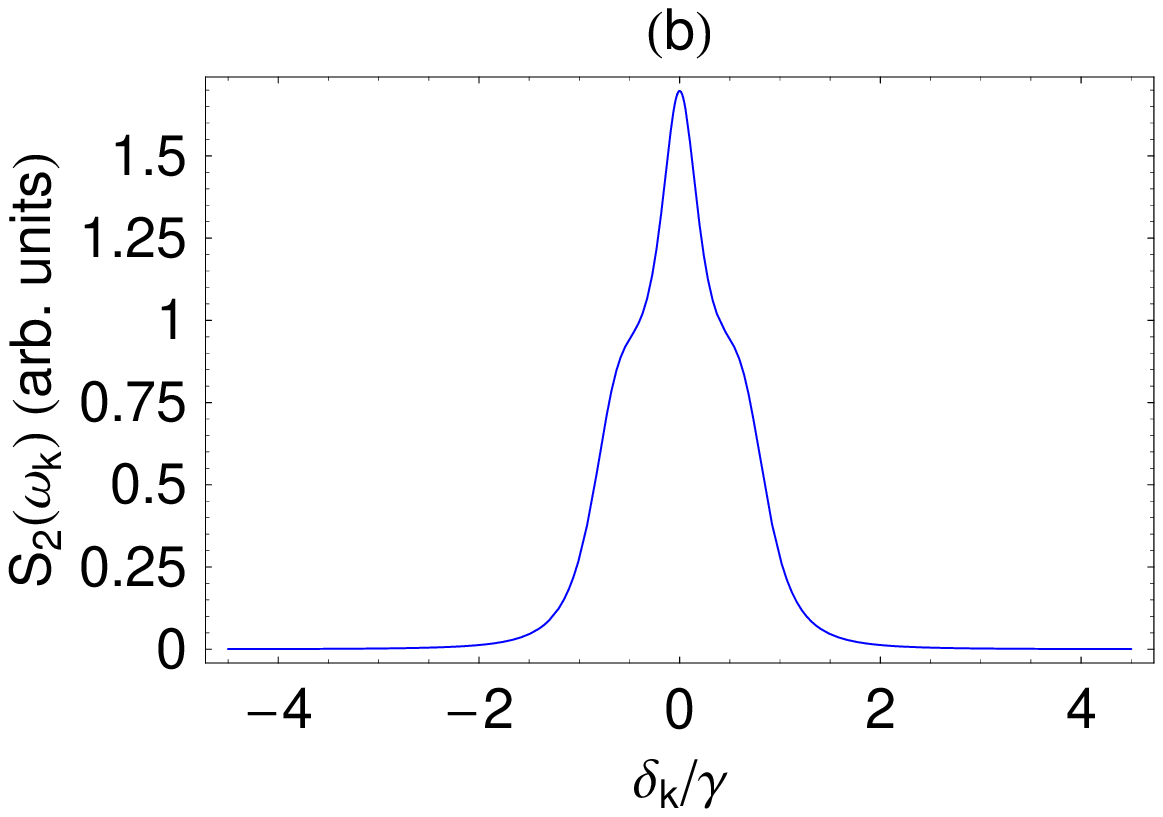}
\includegraphics[width=4 cm]{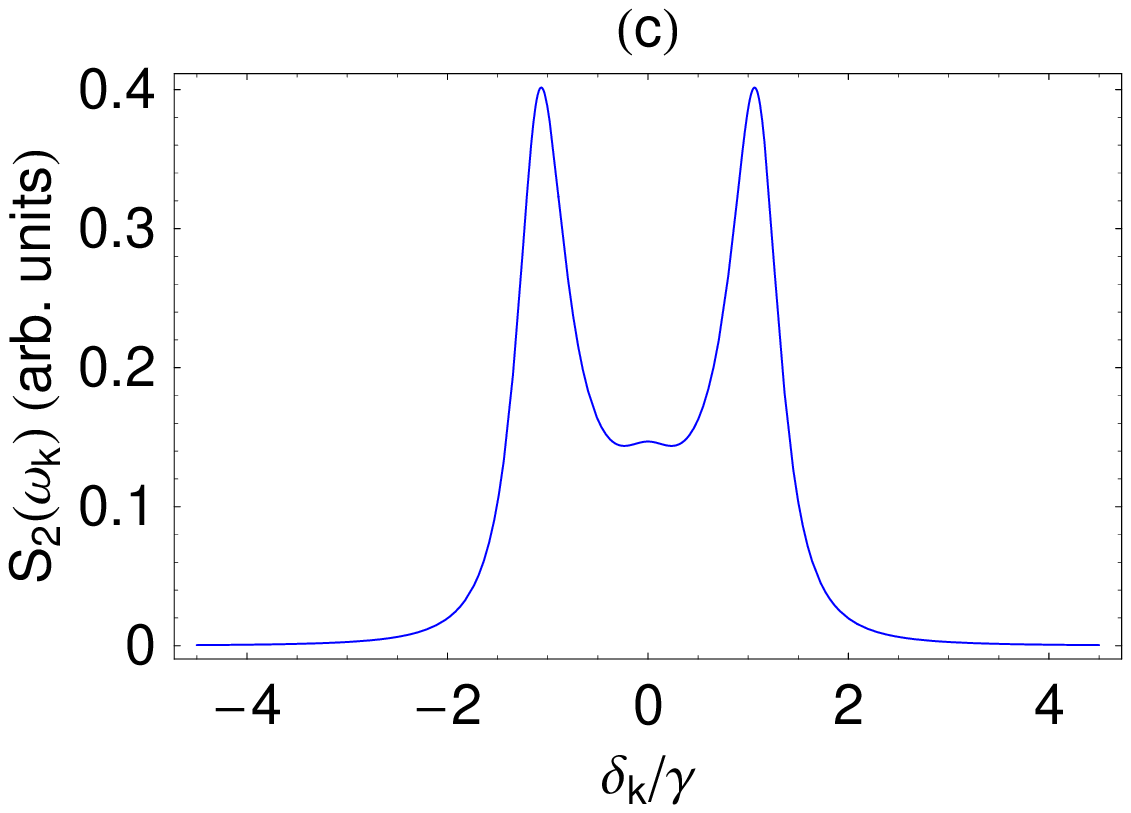}
\includegraphics[width=4 cm]{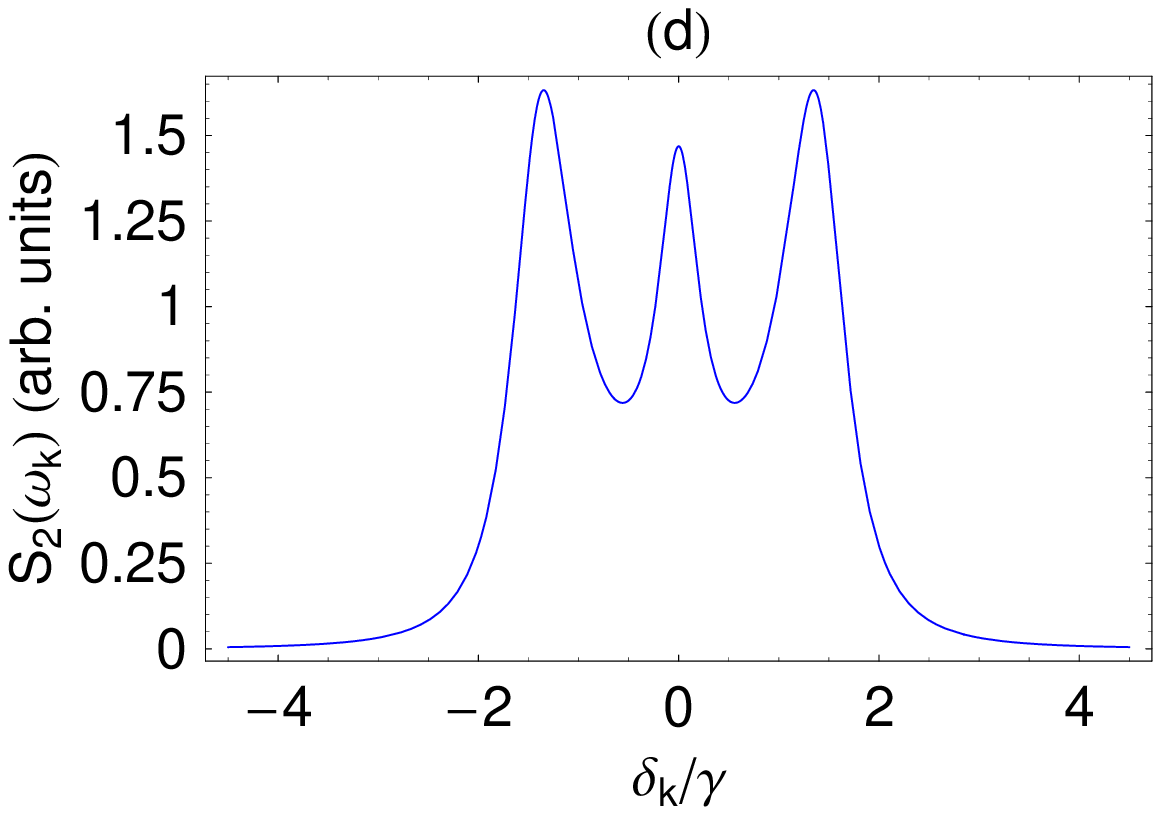}
\includegraphics[width=4 cm]{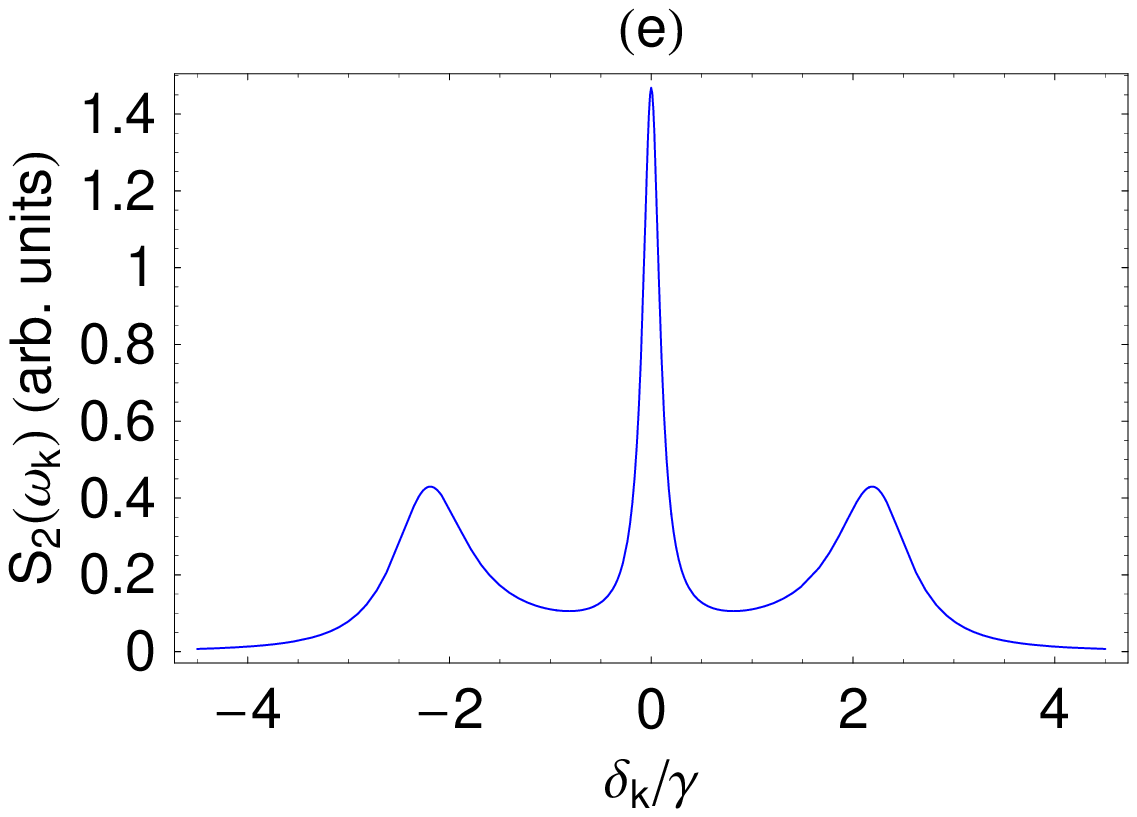}
\includegraphics[width=4 cm]{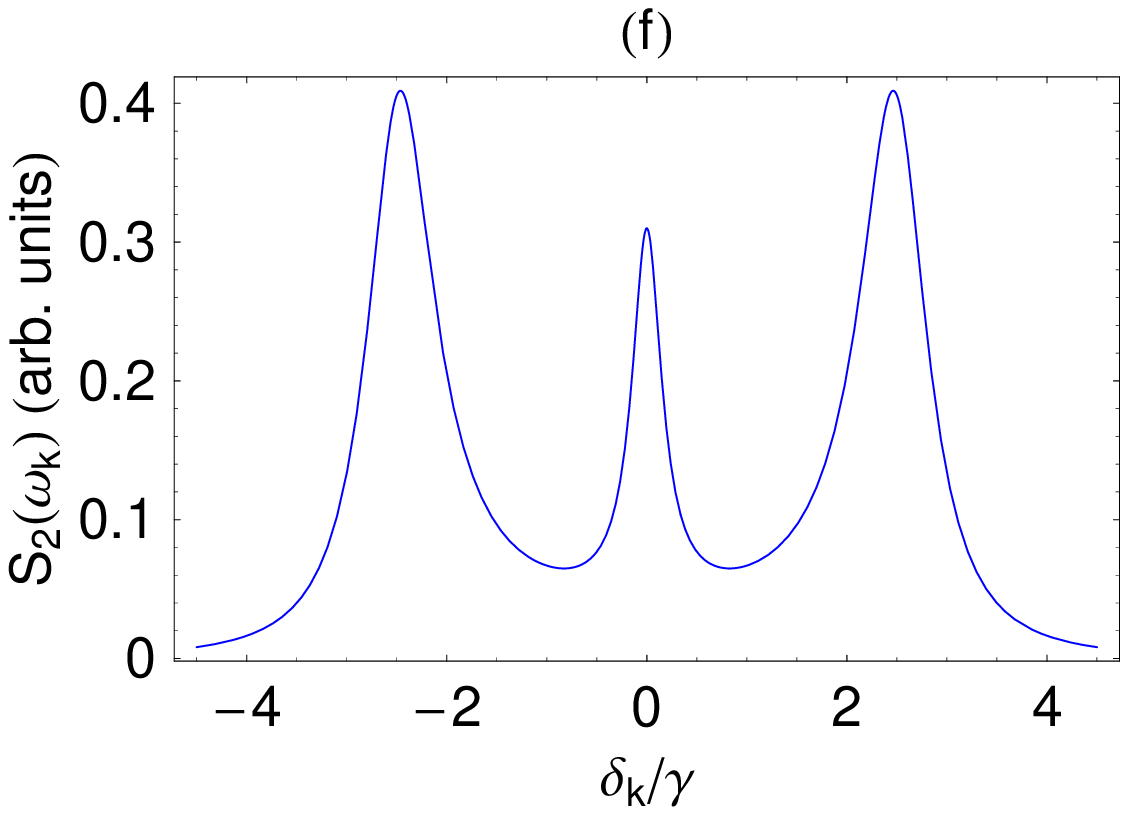}
\caption{(Color online) The spontaneous emission spectra $S_2(\omega_k)$ of level
$|2\rangle$ for the atom is initially prepared in one of the
ground states $|1\rangle$, $a_{1}(0)=1$.
The parameters for the calculations are $\delta_{1}=\delta_{2}=\delta_{3}=0$,
$\gamma_{2}=\gamma_{3}=\gamma$, $\gamma=6.0$ MHz, and $\Omega_{12}=0.5\gamma$.
(a) $\Omega_{24}=\Omega_{23}=0.50\gamma$.
(b) $\Omega_{24}=\Omega_{23}=1.0\gamma$.
(c) $\Omega_{24}=2.0\gamma$, $\Omega_{23}=1.0\gamma$.
(d) $\Omega_{24}=2.0\gamma$, $\Omega_{23}=2.0\gamma$.
(e) $\Omega_{24}=2.0\gamma$, $\Omega_{23}=4.0\gamma$.
(f) $\Omega_{24}=3.0\gamma$, $\Omega_{23}=4.0\gamma$.}
\end{figure}
To find the role of each coupling laser, for a given pump laser L1, we keep the Rabi frequency of the upper coupling laser L3, and only increase the Rabi frequency of laser L2, then the spectrum displays a double-peak structure. The spectrum is similar to a $\Lambda$ coupling scheme as long as the Rabi frequency of L2 is much larger than that of L3, but with a little bump at resonance frequency ($\delta_k=0$) as shown in Fig. 3(c). If we keep the Rabi frequency of laser L2 but increase the Rabi frequency of the upper coupling laser L3 the central component emerges(Fig. 3(d)). As we further increase the strength of Laser L3, the central component is enhanced and its linewidth decreases, while two sidebands are suppressed and separates more with respect to the central component as shown in Fig. 3(e). When we keep the Rabi frequency of the upper coupling laser L3 as the same as in Fig. 3(e) but increase the Rabi frequency of laser L2 the sidebands will be enhanced and the central component is suppressed and the linewidth is broadened as shown in Fig. 3(f). This is quite remarkable, because the spontaneous emission spectral features of level $|2\rangle|$ can easily be controlled by the combination of the Rabi frequencies of two coupling lasers L2 and L3 through the quantum interference effects. The desired frequency component can be enhanced or suppressed, narrowed or broadened. In fact, the linewidth of the central component can be subnatural by adjusting the Rabi Frequencies of laser L2 and laser L3 to have $y_{+}+y_{-}\approx\frac{\gamma_2+\gamma_3}{6}$ according to Eq. (26b). From the experiment point of view, controlling the laser intensity therefore the Rabi frequency of the laser can be easily achieved.
\subsection{(C) Effects of the Decay Rate of the Upper Level}
Above analysis shows that the central component of the spectrum is due to the upper coupling laser L3. To illustrate this we plot two spectra to compare the spectral features for the $\Lambda$ coupling scheme($|1\rangle$-$|2\rangle$-$|4\rangle$, without L3) with the current inverted Y-type scheme in Fig. 4. One may intuitively (but mistakenly) think that the sidebands are two Rabi splitting (Autler-Townes) components due to laser 2 or laser 3. Though there are AT splitting when laser L2 and L3 are strong, the spectrum illustrated here is not an AT splitting of the excitation spectrum which has been reported in our recent experiment~\cite{JCP124.084308.2006}, where the fluorescence of the excited states was detected by scanning the probe laser L1 while the laser L2 and laser L3 were tuned at the corresponding resonance transitions. It is easy to check this by inspecting the separation of the two sidebands of the spectrum. For example in Fig. 4a, the two sidebands are not separated by the corresponding Rabi frequencies($\Omega_{24}$ or $\Omega_{23}$), but by 2$|\delta_{\lambda}|=|\frac{\sqrt{3}}{2}(y_{+}-y_{-})|$ according to Eq. (26a). By inspecting the spectral linewidth of the central component with respect to the decay rate of the upper level $|3\rangle$ $\gamma_3$, we find that the linewidth decreases as $\gamma_3$ decreases for a given Rabi frequency of laser L3 as we show two examples in Fig. 5(a)-(b). The origin of this result is not so obvious by checking Eq. (26b), but if we think about the extreme case: when the upper state is not decaying the coupling  between level $|2\rangle$ to level $|3\rangle$ would be equivalent to a situation that the laser L3 coupling $|2\rangle$ to another ground state level $|3\rangle$. This can be checked in Eq. (16) by setting $\gamma_3=0$ as shown as the dashing lines in Fig. 5(b).
\begin{figure}
\centering
    \includegraphics[width=8 cm]{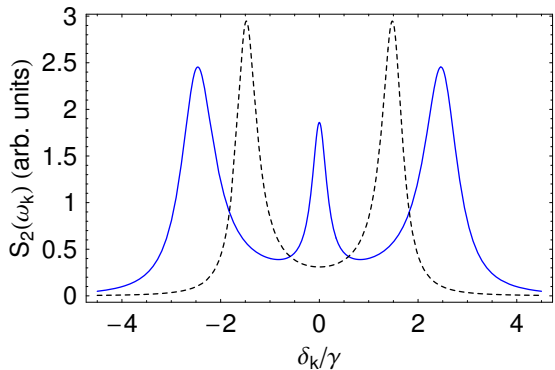}
\caption{(Color online) A comparison of the spontaneous emission of level $|2\rangle$ with a Lambda scheme.
The solid line is for the inverted Y-type scheme with $\Omega_{23}=4.0\gamma$, $\gamma_{3}=\gamma$ and the
dashing lines are for the $\Lambda$ scheme($|1\rangle-|2\rangle-|4\rangle$).
other parameters for the calculations are $\delta_{1}=\delta_{2}=\delta_{3}=0$,
$\gamma_{2}=\gamma=6.0 MHz$, $a_{1}(0)=1, a_{4}(0)=a_3(0)=a_2(0)=0$,
$\Omega_{12}=0.5\gamma$, and $\Omega_{24}=3.0\gamma$.}
\end{figure}
This result can be useful for measuring the decay rate of the upper state. For example, in the case of direct fluorescence detection of $|3\rangle\rightarrow|e\rangle$ transition being not convenient, one can detect the fluorescence of a convenient transition $|2\rangle\rightarrow|g\rangle$ and then  calculate the corresponding $\gamma_3$ by fitting the experimental spectrum using Eq. (26b).
\begin{figure}
\centering
    \includegraphics[width=8 cm]{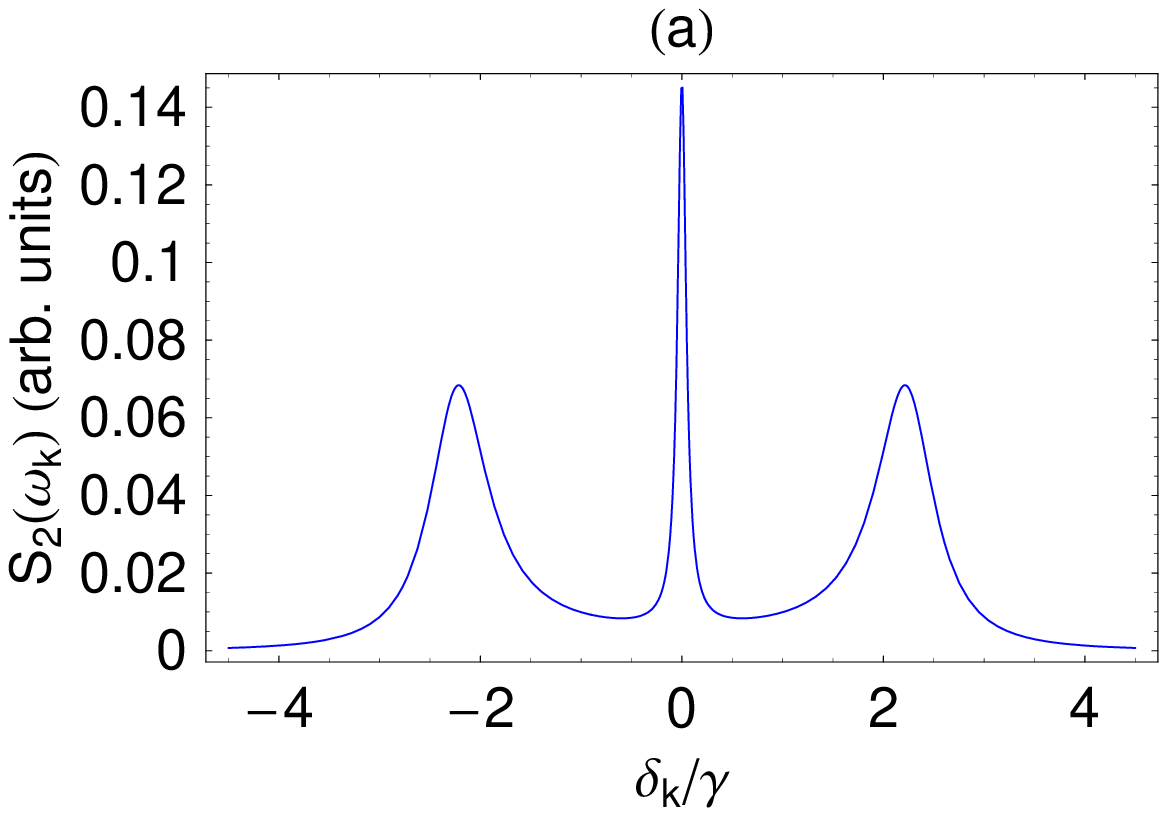}
    \includegraphics[width=8 cm]{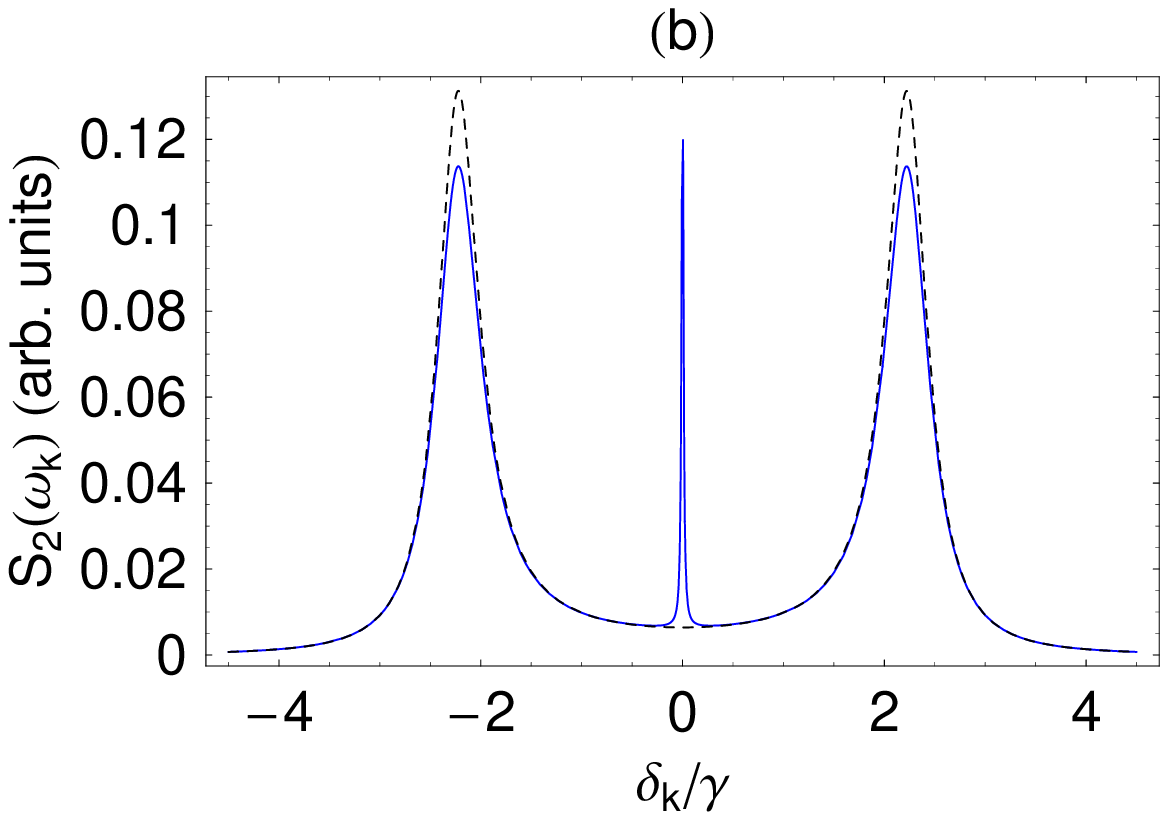}
\caption{(Color online) The dependence of the spontaneous emission spectrum $S_2(\omega_k)$
upon the decay rate of the upper state $|3\rangle$.
(a)$\gamma_{3}=0.5\gamma$ (b)$\gamma_{3}=0.1\gamma$ for the solid line, $\gamma_3$=0 for the dashing lines.
Other parameters for the calculations are $\delta_{1}=\delta_{2}=\delta_{3}=0$;
$\gamma_{2}=\gamma$ and $\gamma=6.0$ MHz; $a_{1}(0)=1$ and $a_{4}(0)=a_3(0)=a_2(0)=0$,
$\Omega_{12}=0.5\gamma$, $\Omega_{24}=2.0\gamma$ and $\Omega_{23}=4.0\gamma$.}
\end{figure}
\section{IV. Summary and Conclusion}
We have investigated the spontaneous emission for an inverted Y-type atom driven by three coherent fields. A wave function approach is used to derive an analytical expression of the spontaneous emission spectrum. We show that quantum interference leads to the spectral line narrowing, spectrum splitting and dark fluorescence. The origin of the spectral characteristics can be explained by the analytical expression with the corresponding physical parameters. We find that the number of spectral components, the spectral linewidth, and relative heights can be controlled by the amplitudes of the coupling fields and the preparation of the initial quantum state of the atom. The numerical results have been presented based on the theoretical model and the role of each parameters is examined. The limitations of this approach are that it may not be able to treat the situation that there is an incoherent pumping among the levels, such as repopulation terms. In these cases a density matrix approach with use of the quantum regression theorem has to be used as in references [15, 42]. The results obtained in this paper do not include the motion of the atom, therefore only valid for Doppler free environments. For a realistic experimental observation and to eliminate Doppler effect, we propose an ultracold atomic system, such as a $^{87}Rb$ for experimental observation since Doppler effect can be negligible in an ultracold system.
\section{Acknowledgements}
This work is supported by the RDG grant from Penn State University.

\end{document}